\documentclass[traditabstract]{aa} 
\usepackage{graphicx}
\usepackage{txfonts}
\begin{document}
\title{More on molecular excitations: Dark matter detection in ice}

\author{J.Va'vra
\inst{}
}

\institute{SLAC, Stanford University, CA94309, U.S.A.\\
\email{jjv@slac.stanford.edu}
}

\date{Received May 30, 2016}


\abstract
{In this paper we investigate di-atomic molecules embedded in ice crystals under strain. In this environment coherent vibrations of many OH-bonds may be generated by one WIMP collision. The detection of such multiple-photon signals may provide a signature of a 100~GeV/c${^2}$ WIMP. To do a proper lab test of “WIMP-induced” multi-photon emission is very difficult. We suggest that Ice Cube make a search for multi-photon events, and investigate whether the rate of such events exhibits yearly modulation.} 


\maketitle
%

\section{Introduction}

A recent paper [\cite{Vavra_2015}] discussed the possibility of exciting molecular vibrations of di-atomic molecules when a nucleus, either oxygen or hydrogen, is struck by a Dark Matter (DM) particle. An example of such a molecule is H$_{2}$O, which has 3 vibrational modes involving the OH-bond (two are stretching and one is bending, or scissoring). Such vibration leads to single-photon emission, mostly in the IR wavelength region, or to the emission of heat; the emission of visible wavelength photons is suppressed by at least 4-6 orders of magnitude. 

\begin{figure}[tbp]
\includegraphics[width=0.45\textwidth]{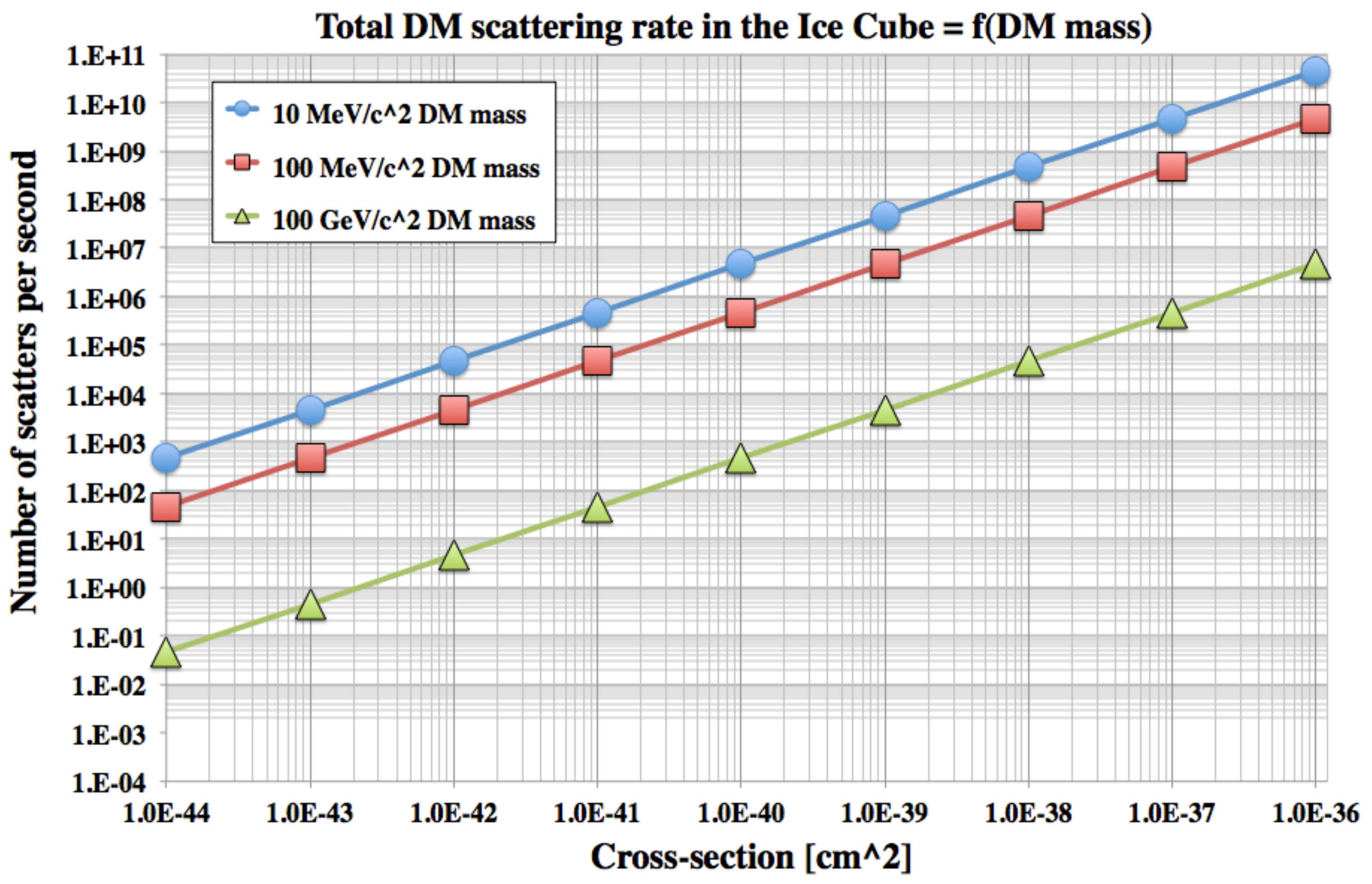}
\caption{Expected Ice Cube experiment interaction rate as a function of WIMP-nucleus cross section for different WIMP mass values and assuming that the WIMP has a velocity of 500 km/sec relative to Earth.}
\label{fig:Ice_cube_rate}
\end{figure}

Assuming that the DM-nucleus scattering cross section is $\sim$10$^{-44}$~cm$^{2}$ or even smaller, a very large target is needed. An example of one such detector is the Ice Cube experiment, with its huge active volume (1000~m~x~1000~m~x~1000~m), large detection system of 5160 PMTs, and ice with its large photon attenuation length of more than $\sim$100~meters. Figure~\ref{fig:Ice_cube_rate} shows the expected Ice Cube experiment interaction rate as a function of WIMP-nucleus cross section; it is huge. However, a simple calculation shows that it is impossible to detect DM with a sufficiently high S/N ratio using the present PMTs,\footnote[1]{Average Ice Cube PMT noise is $\sim$280~Hz/PMT.} if we assume a single photon is produced per WIMP collision.

However, ice under extreme pressure is a much more complicated substance than the OH-molecular system. As discussed in this note, coherent effects may lead to multi-photon emission, and so we consider what happens if a 100~GeV/c${^2}$ WIMP mass hits either the oxygen nucleus or the proton embedded in the strained crystal structure of ice.

\section{The crystal structure of ice}

\begin{figure}[tbp]
\includegraphics[width=0.45\textwidth]{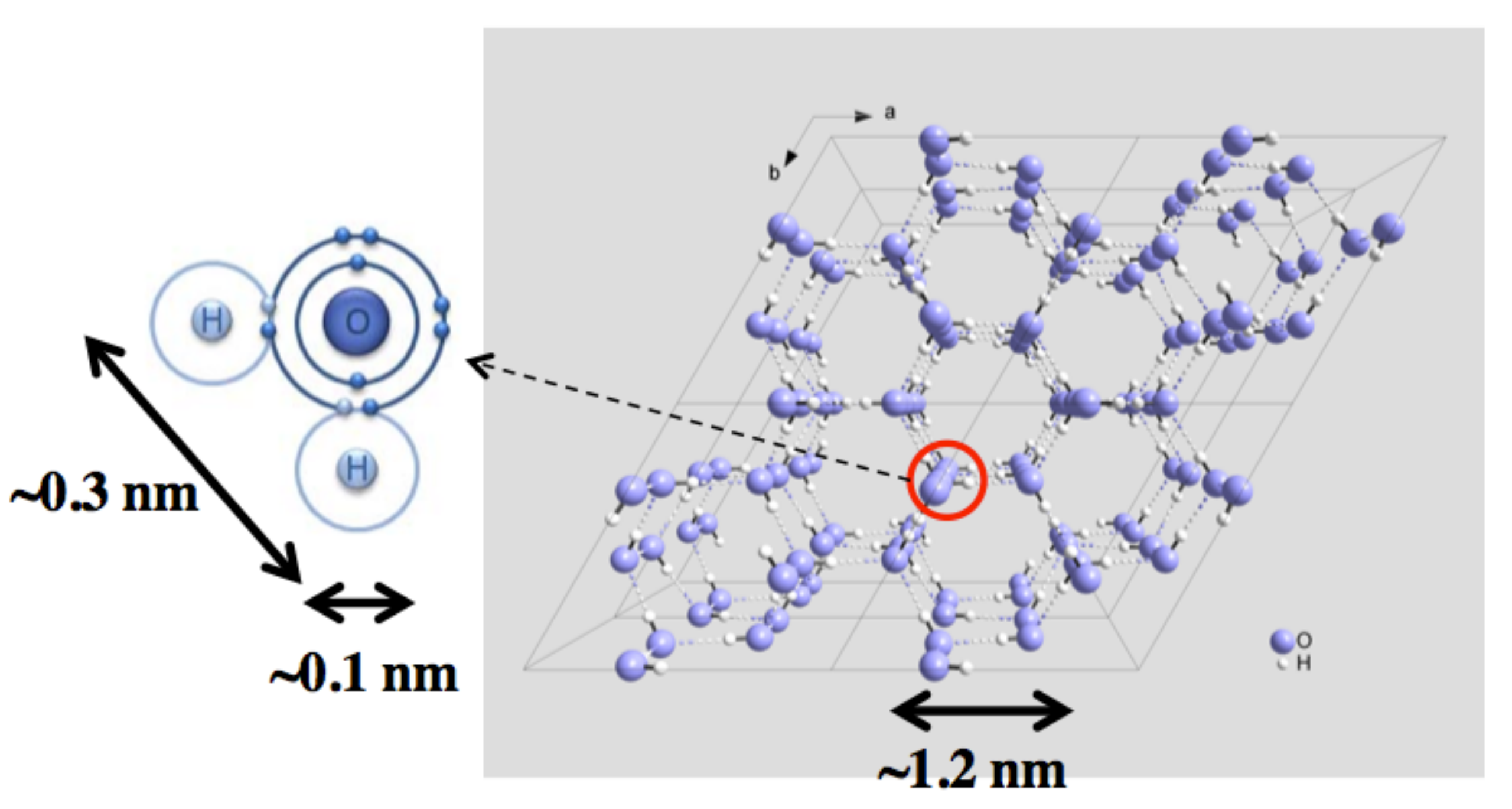}
\caption{Hexagonal ice crystal structure Ih, found in most locations on Earth. There are many OH-bonds within each hexagon. One can see the approximate sizes of the hexagon, a water molecule, and a hydrogen atom. From the point of view of this paper, this is similar to a system of many springs.}
\label{fig:Ice_Ih}
\end{figure}

\begin{figure}[tbp]
\includegraphics[width=0.4\textwidth]{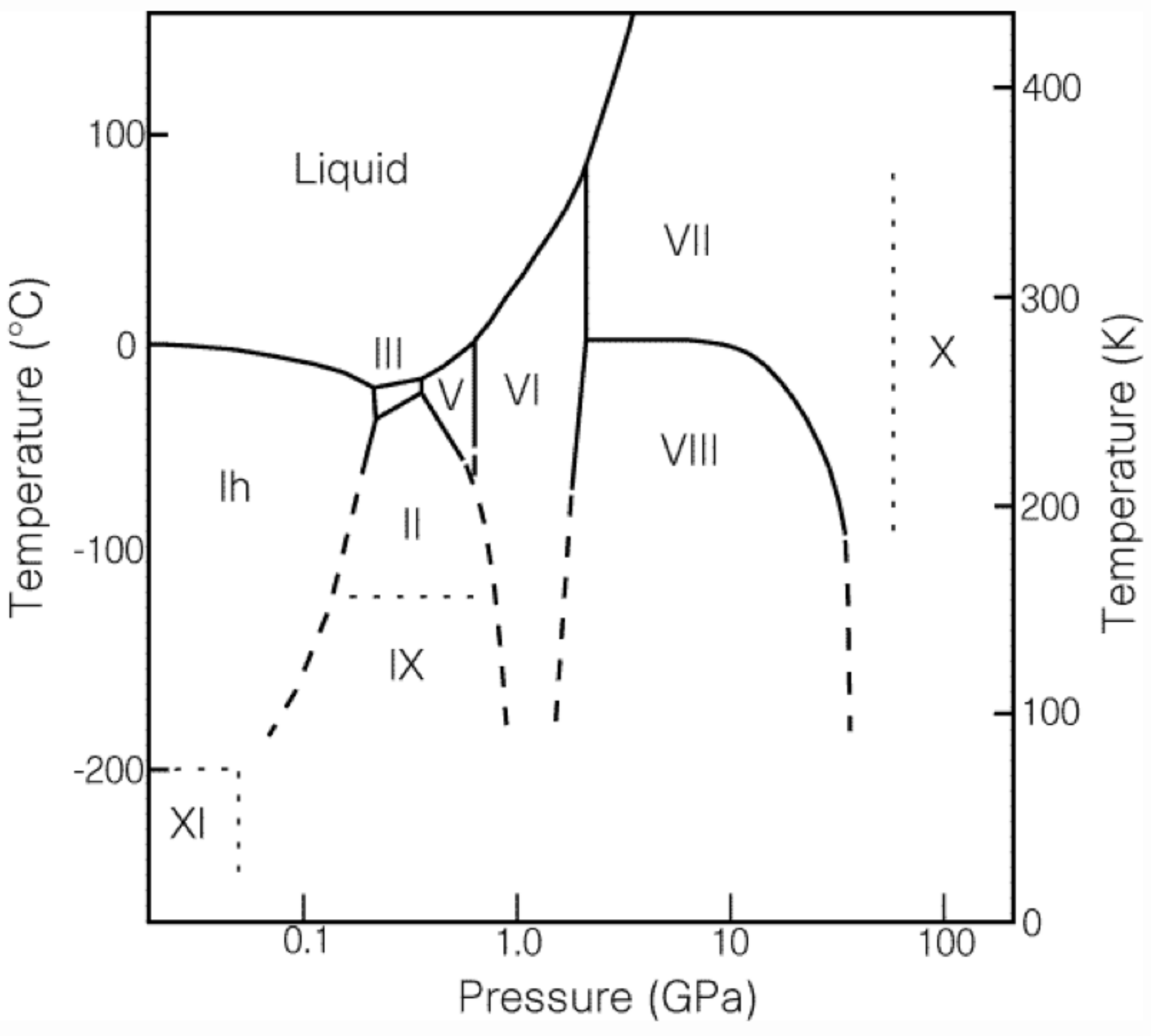}
\caption{Ice phases as a function of pressure and temperature [\cite{Lobban_1998}]. Ice crystal structure type Ih is found in most locations on Earth. The ice type in the area of the Ice Cube experiment is believed to be Ih, stressed to $\sim$23~MPa, although we were not able to find an experimental proof of its pressure. As ice pressure increases, the type changes.}
\label{fig:Ice_phases}
\end{figure}

\begin{figure}[tbp]
\includegraphics[width=0.4\textwidth]{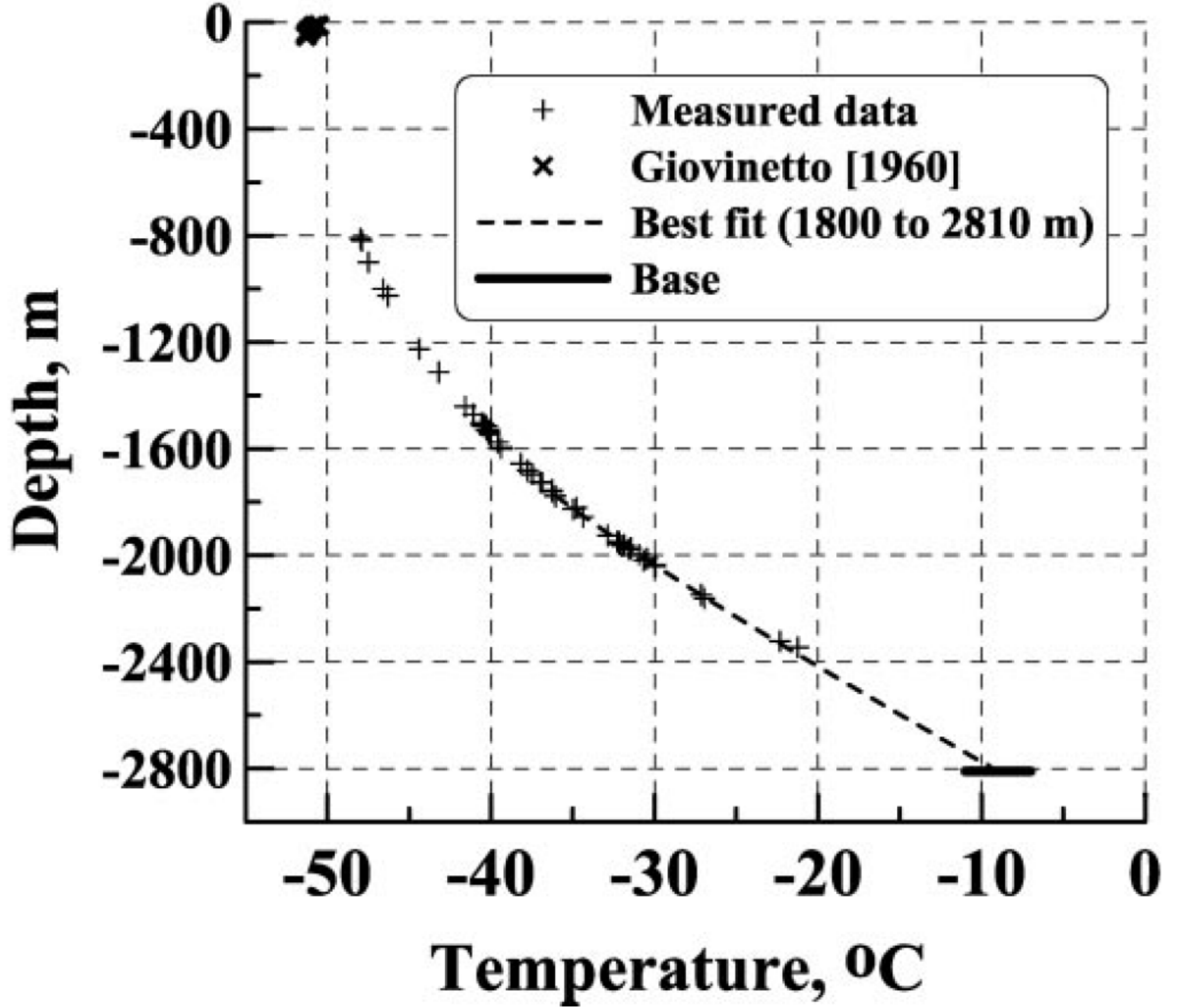}
\caption{Ice temperature measured in Amanda boreholes as a function of depth [\cite{Price_2002}].}
\label{fig:Ice_temperature}
\end{figure}

  Ice has very rich molecular structure, with at least fifteen forms observed experimentally. Figure~\ref{fig:Ice_Ih} shows the hexagonal ice crystal structure type Ih, found in most locations on Earth. Figure~\ref{fig:Ice_phases} shows its possible phases as a function of temperature and pressure [\cite{Lobban_1998}]; we see that as long as the ice pressure is below $\sim$0.2~GPa, it remains type Ih. Figure~\ref{fig:Ice_temperature} shows the ice temperature as a function of depth, as measured in Amanda near the South Pole [\cite{Price_2002}]. We can see that the ice temperature increases with pressure; it approaches -20$^0$C at a depth of $\sim$2500 meters.\footnote[2]{The Amanda experiment quotes ice temperatures between -43$^0$C at 1450~meter depth and -20$^0$C at 2450~m depth [\cite{Abbasi_2011}].} The pressure approaches $\sim$230~bars, or $\sim$23~MPa, for ice $\sim$2500 meters below the surface, based on gravitational force alone. Therefore we might expect, based on the phase diagram of Fig.~\ref{fig:Ice_phases}, that the ice lattice configuration remains type Ih for ice $\sim$2500 meters below the surface. However, we point out that we are neglecting possible stresses due to slow horizontal ice movements over uneven bedrock shape\footnote[3]{The Antarctic ice is actually moving very likely non-uniformly. Near the South Pole, ice slowed down because high mountains underneath are blocking its movements. Nevertheless, ice under the Ice Cube experiment still moves slightly as measurements with inclinometers, located below a depth of ~2800 meters, indicate. Therefore there is a pressure buildup on this ice region, both from sides and from upward forces driven by a shape of the bedrock.}, or long-term effects over geological time of ice under stress, effects of long-term bombardment by cosmic rays, etc. 

Whatever ice-type Ice Cube has at a depth of $\sim$2500 meters, it is reasonably uniform, judging from the long light and sound attenuation lengths observed (the sound attenuation length of ice near the Ice Cube experiment is more than $\sim$300~meters [\cite{Tosi_2010}], and the light attenuation length is $\sim$110~meters on average). This indicates a great degree of crystal uniformity, and only a small presence of dust or voids. 

It is also known that ice under stress can yield triboluminescence.\footnote[4]{Triboluminescence is an optical phenomenon in which light is generated through the breaking of chemical bonds in a material when it is pulled apart, ripped, scratched, crushed, or rubbed. Although it is known phenomenon, it has not been studied quantitatively in detail.} For example, ice in the Ice Cube detector holes, 2500~meters below the surface, produced triboluminescence after it was subject to large stresses due to ice expansion (detector enclosures were designed to withstand pressure up to $\sim$690~bars); it was observed that the triboluminescence light decays exponentially over several years. Finally, bursts of triboluminescence photons ($\sim$200 counts/sec lasting over a second) were observed when 5ml blocks of purified H$_{2}$O ice (Ih) at temperatures just below 0$^0$C were dropped into liquid nitrogen while viewed by a photon counter [\cite{Quickenden_1998}].

\section{Maximum nuclear recoil energy}

\begin{figure}[tbp]
\includegraphics[width=0.45\textwidth]{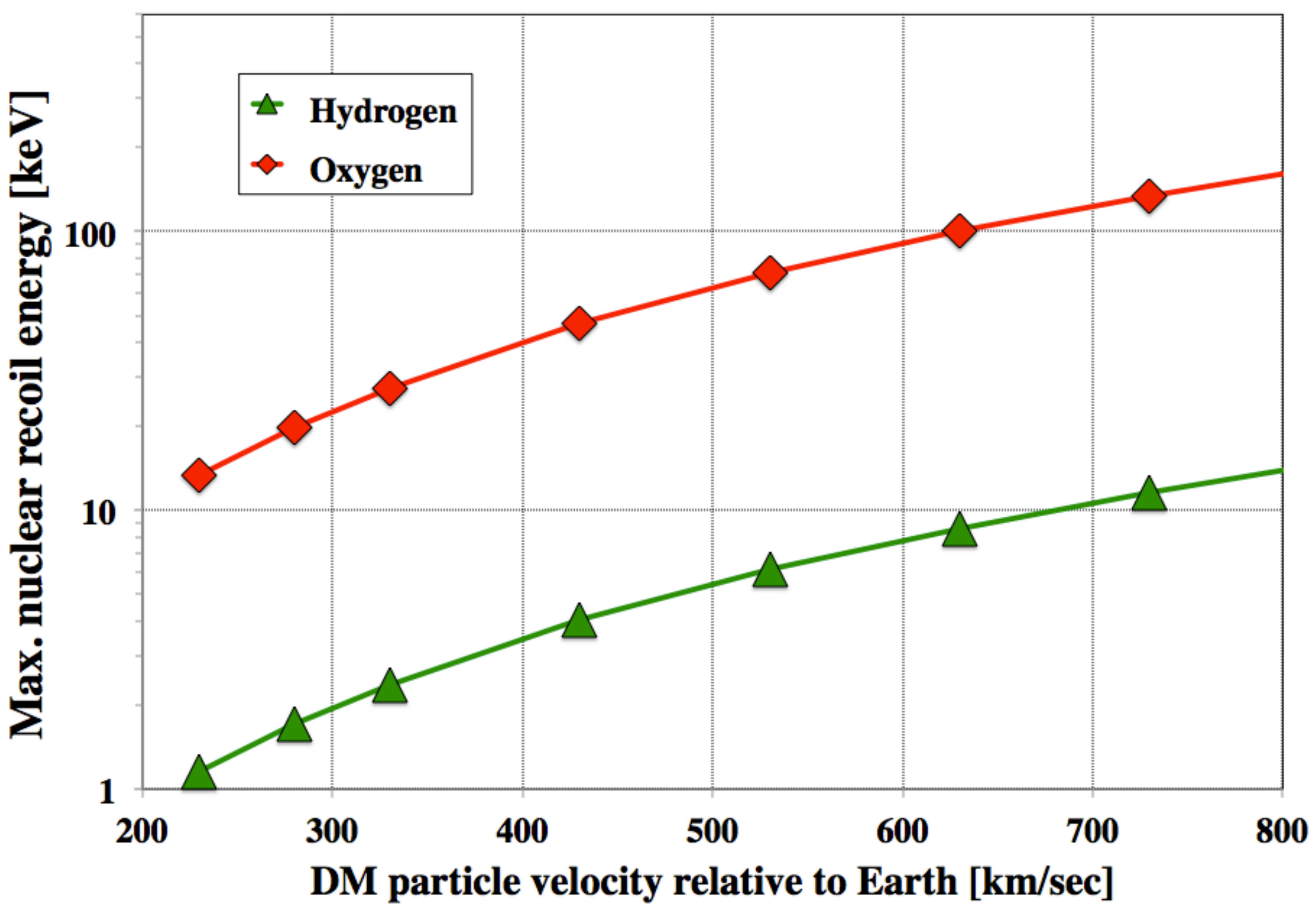}
\caption{Maximum recoil energy of an oxygen or hydrogen nucleus struck by a 100~GeV/c${^2}$ DM particle as a function of its relative velocity to Earth.}
\label{fig:Max_recoil}
\end{figure}

   Figure~\ref{fig:Max_recoil} shows a maximum recoil energy of $\sim$60~keV for a 100~GeV/c${^2}$ WIMP traveling with a velocity of $\sim$500~km/sec striking an oxygen nucleus, and  $\sim$8~keV if it strikes a hydrogen nucleus. Such an impact would certainly break off the oxygen or hydrogen nucleus from its molecular bond in the hexagonal lattice. This would upset the local electrostatic balance in one single hexagon, thus leading to its deformation.  

\section{Calibration of photon yield in ice}

In this section we argue that we do not have a relevant calibration test, which would enable us to predict the photon yield per unit of energy deposited by a WIMP-like particle.

\begin{figure}[tbp]
\includegraphics[width=0.45\textwidth]{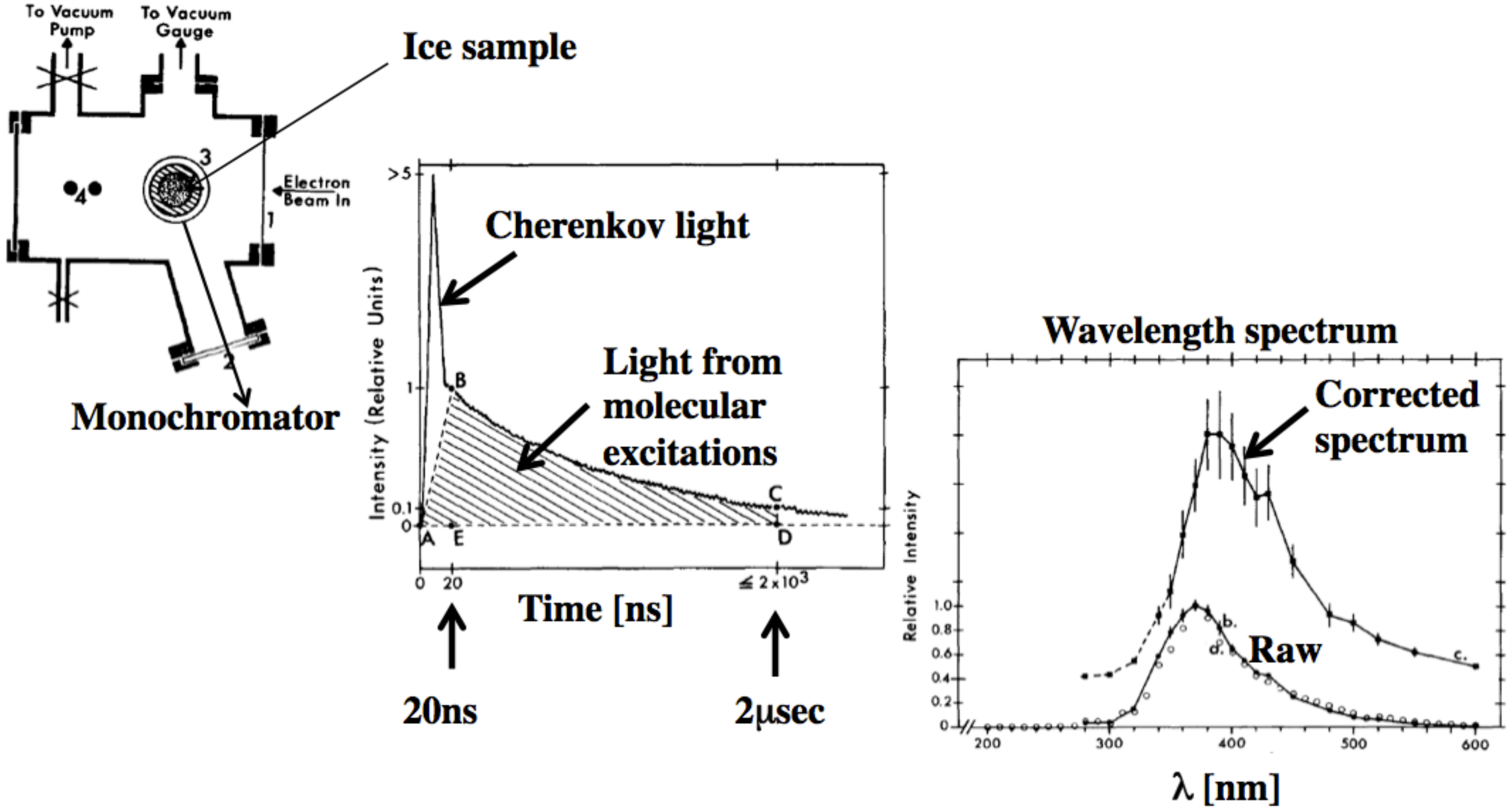}
\caption{Principle of calibration test [\cite{Quickenden_1982}] measuring photon luminescence yield per unit of deposited energy. The test used a $\sim$0.5~MeV electron beam and ice at 88$^0$K. We see a prompt Cherenkov peak followed by molecular vibration luminescence, which is mostly in the visible wavelength region.}
\label{fig:Qickenden_calibration}
\end{figure}

The only calibration of which we are aware are results from two references, one used a continuous UV photon beam [\cite{Quickenden_1985}], and the other used a pulsed $\sim$0.5 MeV electron beam [\cite{Quickenden_1982}]. Figure~\ref{fig:Qickenden_calibration} shows the result from a pulsed electron beam striking ice held at a temperature of 88$^0$K. We see a prompt Cherenkov peak followed by molecular vibration luminescence, which is mostly in the visible wavelength region. The measured average photon yield was $\sim$2.4x10$^{-6}$~photons/eV of deposited energy.\footnote[5]{Specifically the emission was observed in the three wavelength regions, 280-300~nm ($\sim$10$^{-9}$~photons/eV), 340-460~nm ($\sim$2x10$^{-6}$~photons/eV), and 500-600~nm ($\sim$10$^{-7}$~photons/eV).} This emission is caused by "normal" light emission from electromagnetic deposition of energy in unstrained ice. This means that a $\sim$50~keV nuclear recoil will produce $\sim$0.12 visible wavelength photons on average, and therefore we would be dealing with single photon emission only per WIMP collision.\footnote[6]{One should note that the Quickenden reference [\cite{Quickenden_1982}] did not investigate a possible multiple-photon emission, which would be difficult to do in their simple setup; such test would need a 4$\pi$ detection geometry.} That would mean that the signal-to-noise ratio in the Ice Cube experiment would be too small to detect an annual modulation due to variation of WIMP velocity relative to Earth.

  However, there are three problems with this calibration. First, it deals with ice produced in the lab, and not with strained ice, as in the Ice Cube experiment. The second argument is that De~Broglie wavelength of $\sim$0.5 MeV electrons is $\sim$5.5~nm, which is about 5-times larger than the ice hexagon in the ice type Ih, and is much larger than the oxygen diameter of the nucleus. In fact, we would argue that a $\sim$0.5~MeV electron does not see oxygen nuclei at all. This means that such electrons interact with the hexagonal and OH structure differently\footnote[7]{We have not calculated this difference.} than would a 100~GeV/c${^2}$ WIMP which strikes the oxygen nucleus directly, as its De~Broglie wavelength of $\sim$7.9~fm is comparable to the size of the nucleus (the diameter of the oxygen nucleus is $\sim$8.8~fm), and hence the third problem is that this calibration does not use a particle beam which properly simulates the effect of a 100~GeV/c${^2}$ WIMP. 

\section{Model of DM-induced Triboluminescence producing multi-photon events}

   We would argue that a crystal structure under stress will behave differently than unstrained ice. When the oxygen nucleus is knocked out of its location, the electrostatic balance within the hexagon is upset, it breaks locally under large external pressure, causing OH-vibrations (see Fig.~\ref{fig:Ice_Ih} for locations of OH-bonds within each hexagon). External stress will help the "rupture", and will provide additional energy to the nuclear recoil; it acts as a "nature-provided" amplifier. This rupture will propagate to neighboring hexagons, causing subsequent OH-vibrations. This wave will propagate with the velocity of sound (3.92~$\mu$m/nsec [\cite{Price_2006}]) for some distance. It is a form of DM-triggered triboluminescence. Such behavior was not tested in the Quickenden calibration.

  If the WIMP mass is much smaller, the advantage derived from stressed ice crystal structure will be smaller. For example, a $\sim$1~MeV/c${^2}$ WIMP has a De~Broglie wavelength of $\sim$0.79~nm, which is comparable to the size of the ice hexagon, and so each DM particle will not see individual nuclei. Also the nuclear recoil energy would be too small to be detected though this mechanism; instead the DM particle will interact will shell-electrons in the entire crystal structure. We would expect that the Quickenden calibration would then be more relevant, and so there would be no possibility of detecting a small DM mass with the present Ice Cube PMT detectors.

\begin{figure}[tbp]
\includegraphics[width=0.40\textwidth]{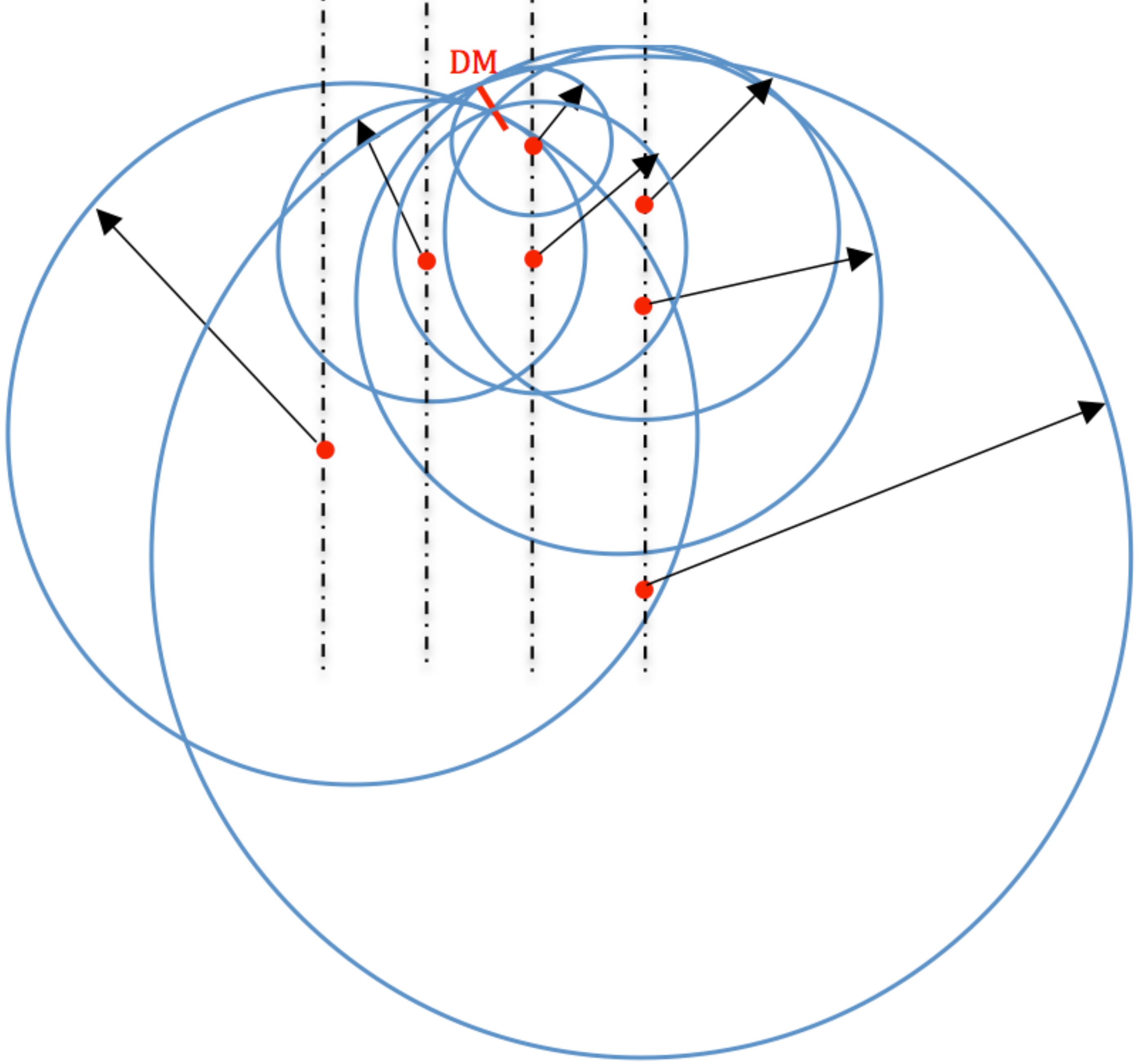}
\caption{One could attempt to find DM hit location in the Ice Cube for some photon bursts is by finding an intercept of 3 spheres, each having a radius of time-of-propagation of light in ice, after a common time offset is subtracted. Obviously one has to have at least 3 detected photons produced in one location for this to work. In this example we assume two bursts, which are separated by a long time delay due to a slow propagation of sound impulse in ice.}
\label{fig:DM_vertex}
\end{figure}

   If the WIMP-induced triboluminescence occurs in the Ice Cube, one could search for locations of these bursts as indicated on Fig.~\ref{fig:DM_vertex}, provided that at least 3 photons are detected corresponding to the same localized burst. Can one detect a DM direction for events where two different 3-photon bursts along a single "rupture" are detected ?
   
\section{Conclusion}

  Stressed ice near the South Pole at a depth of 2500~meter may produce multi-photon emission if an oxygen nucleus is struck by a 100~GeV/c${^2}$ WIMP and removed from the strained ice lattice structure. When the oxygen nucleus is displaced from its location, the electrostatic balance within the hexagon is upset, the hexagon structure breaks locally under large external pressure, causing many OH-vibrations. External stress will amplify this lattice rupture and provide an additional contribution to the primary nuclear recoil energy. This rupture wave will propagate with the velocity of sound to neighboring hexagons, causing many subsequent OH-vibrations. It is thus a form of DM-triggered triboluminescence. If the multi-photon emission does occur, the S/N ratio for DM detection in the Ice Cube experiment will be significantly enhanced.

  To do a proper lab test of “WIMP-induced” multi-photon emission is very difficult. We suggest that Ice Cube make a search for multi-photon events, and investigate whether the rate of such events exhibits yearly modulation.

\section{Acknowledgements}

 I thank Prof. Francis Halzen, Dr. Kael Hanson, Dr. Blair Ratcliff and Dr. Bill Dunwoodie for several useful comments.


\begin{thebibliography}{}

\bibitem{Vavra_2015} [1] J.Va'vra, "Molecular excitations: a new way to detect Dark matter", Physics Letters B 736, 169 (2014).
\bibitem{Tosi_2010}  [2] Delia Tosi, "Measurement of Acoustic Attenuation in South Pole Ice with a Retrievable Transmitter", Dissertation, Humboldt University, Berlin (2010), 
and R. Abbasi et al. arXiv:1004.1694v2, Nov. 2010.
\bibitem{Lobban_1998} [3] C.Lobban, J.L.Finney and W.F.Kuhs, Nature 391, 268 (1998).
\bibitem{Price_2002} [4] P.B.Price et al., Proc. National Academy Sciencies 99, 7844 (2002). 
\bibitem{Abbasi_2011} [5] R.Abbasi et al., ArXiv:1108.0171v2, Sep. 2011.
\bibitem{Quickenden_1982} [6] T.I.Quickenden, S.M.Trotman and D.F.Sangster, "Pulse radiolytic studies of the ultraviolet and visible emissions from purified ice", J. Chem. Phys. 77, 3790 (1982). 
\bibitem{Quickenden_1985} [7] T.I.Quickenden, S.M.Trotman and D.F.Sangster, "UV excited luminescence from crystalline ice", Chem. Phys. Letters 114, 164 (1985).
\bibitem{Quickenden_1998} [8] T.I.Quickenden, B.J. Selby, C.G. Freeman, "Ice Triboluminescence", J. Phys. Chem. A, 102, 34 (1998).
\bibitem{Price_2006} [9] P.B. Price, "Attenuation of acoustic waves in glacial ice and salt dome", J. of Geophysical Research, 111, B02201 (2006).


\end{thebibliography}
\end{document}